\documentclass[twocolumn,showpacs,aps,prl]{revtex4}

\usepackage{graphicx}%
\usepackage{dcolumn}
\usepackage{amsmath}
\usepackage{latexsym}

\begin {document}

\title
{
Phase transition in a directed traffic flow network
}
\author
{
 G. Mukherjee$^{1,2}$ and S. S. Manna$^1$ 
}
\affiliation
{
$^1$Satyendra Nath Bose National Centre for Basic Sciences
    Block-JD, Sector-III, Salt Lake, Kolkata-700098, India \\
$^2$Bidhan Chandra College, Asansol 713304, Dt. Burdwan, West Bengal, India 
}
\begin{abstract}

      The generic feature of traffic in a network of flowing electronic data packets
   is a phase transition from a stationary free-flow phase to a continuously growing
   congested non-stationary phase. In the most simple network of directed oriented
   square lattice we have been able to observe all crucial features of such
   flow systems having non-trivial critical behavior near the critical point
   of transition. The network here is in the shape of a square lattice and data packets
   are randomly posted with a rate $\rho$ at one side of the lattice. Each packet
   executes a directed diffusive motion towards the opposite boundary where it is
   delivered. Packets accumulated at a particular node form a queue and a maximum
   of $m$ such packets randomly jump out of this node at every time step 
   to its neighbors on a first-in-first-out (FIFO) basis. The phase transition
   occurs at $\rho_c=m$. The distribution of travel times through the system
   is found to have a log-normal behavior and the power-spectrum of the load time-series shows
   $1/f$ like noise similar to the scenario of Internet traffic.

\end{abstract}
\maketitle

\section{Introduction}

      The transport of matter and propagation of information in biological, social and electronic
   communication systems etc. remains significantly important in different branches of physics,
   more generally in natural science since years. Evidently the prime objective is to make the transport
   or communication processes more efficient in these systems. In particular, one aims at maximizing
   the flow at the same time minimizing the delivery time and loss and of course robustness against attack
   and failure. The effect of the local and global
   topological properties of the system and the microscopic dynamic process involved
   with the flow are considered as the two basic ingredients of these complex
   dynamical process.

     Research on highway traffic as a field of applied physics is already decades old.
   Study on information network traffic is comparatively new. It is evident from the
   empirical observation on Internet traffic \cite {Csabai,Takayasu,Leland}
   and vehicular flow \cite {Kerner} in network of highways that both posses
   similarity in many respects.  In highway network it was observed
   that on increasing the vehicle density a well-defined transition  occurs at a
   critical density separating the free-flow phase and the
   jammed phase. At the critical point the jam or congestion occurs as back-propagating waves
   with fractal properties \cite {Gabor}. In the Internet network it is found that the ping-time
   statistics, in which the time taken by a packet to move from source to
   destination and back were measured,
   show critical dynamics and $1/f$ noise spectrum similar to the scenario of vehicular
   traffic \cite {Csabai}.

    Observation of real computer network dynamics also reveals following behavior, namely: \\
i. Distribution of file sizes is log-normal. \\
ii. Inter-arrival times has a power law distribution. \\
iii. Traffic load time series data shows $1/f$ type fluctuation near the critical point \\
iv. `Ping' experiment data shows round-trip time $\tau_L$ distribution has a log-normal behavior.

\begin{figure}[top]
\begin{center}
\includegraphics[width=8.5cm]{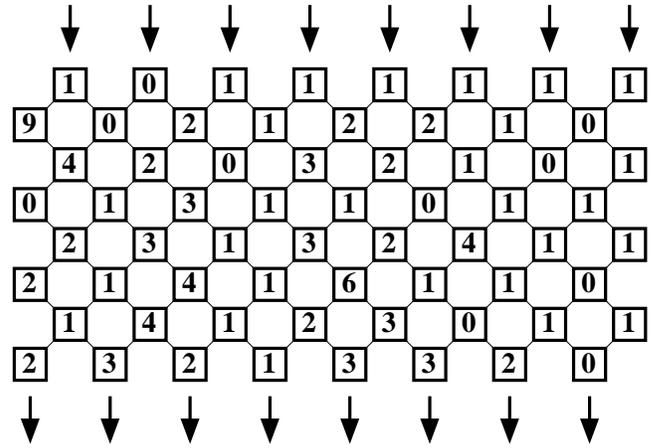}
\end{center}
\caption{
A configuration of the number of data packets at each node of a $8 \times 8$ system
with the posting rate $\rho=0.8$ in the stationary state.
}
\end{figure}

      However the topological structure of the Internet and the highway network are far from being similar.     
   It has been observed  that the nodal degree distributions 
   (degree $k$ of a node is the number of links meeting to it) of 
   the Internet \cite {Faloutsos} and the World-wide web \cite {web} as well as many other real-world
   networks have power law tails: $P(k) \sim k^{-\gamma}$ and  cannot be modeled by
   simple random graphs. This is in contrast to the well known random graphs introduced by Erd\"os and R\'enyi,
   whose degree distribution is Poissonian 
   \cite {Erdos}. Due to the absence of a characteristic value
   for the nodal degree these  new class of networks are called `scale-free networks' (SFN)
   \cite {barabasi, linked,review}. Barab\'asi and Albert (BA) had grown
   scale-free graphs where a fixed number of vertices are added at each time and are
   linked to the growing graph with a linear attachment probability \cite {barabasi}. On the other hand
   topological structure of highway network may show small-world behavior in some cases but
   its degree distribution could not be a power law for practical reasons. 

\begin{figure}[top]
\begin{center}
\includegraphics[width=8.5cm]{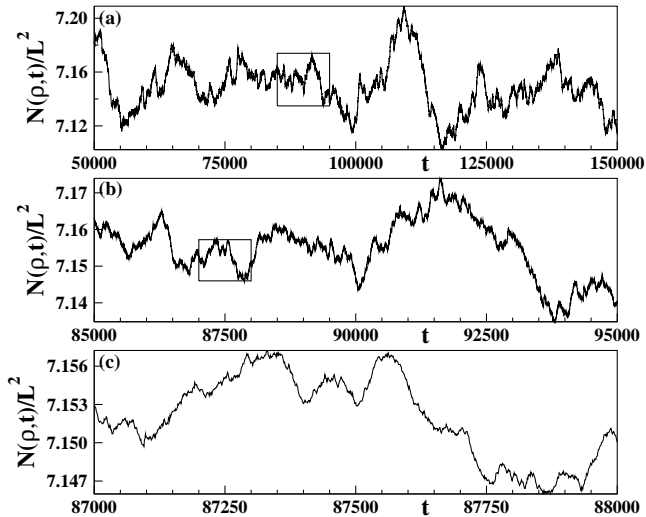}
\end{center}
\caption{
Self-similar fluctuations of the average load per site $N(\rho,t)/L^2$ with time $t$
are displayed.
(a) Average load of a system of size $L=128$ in the stationary state is plotted for the posting rate $\rho = 0.96$ over
a range of 100000 time units.
(b) Magnification of the boxed region in (a) with the horizontal and vertical scale factors
10 and 2.72 respectively.
(c) Magnification of the boxed region in (b) with the horizontal and vertical scale factors
10 and 3.48 respectively.
}
\end{figure}

      So the question is whether the self-similarity and long-range dependence of traffic flow
   and congestion are topological in nature or if they are caused only by the microscopic
   dynamic properties associated with the generation and flow of traffic, such as 
   posting rate, fluctuations in posting rate or routing schemes.

      Traffic system usually involve queues and in the simplest information
   traffic system consisting of a random information input and a buffer shows
   a phase transition behavior when the buffer capacity is infinite \cite
   {Takayasu}. When the mean input rate is smaller compared to maximum possible output rate,
   the average accumulation of information at the buffer is finite and this is called the
   `free' phase. As the mean input rate of information is increased the average
   accumulation at the buffer increases, and at a critical point the averaged
   accumulation diverges. The critical point is defined by the simple condition
   that the mean input rate is equal to the maximum output rate.

      This phase transition behavior is local and can occur in any buffer system
    because of the general nonlinear response of the buffer. But the `ping'-experiment
    indicates phase transition of the whole network due to propagation of
    congestion among jammed nodes and shows $1/f$ fluctuation at the critical point \cite {Takayasu,Leland}.
    The whole system was also considered as a vast ensemble of `phase transition elements'
    and the system properties are outcome of the interactions between these individual
    `elements'  \cite {Takayasu1}.
\begin{figure}[top]
\begin{center}
\includegraphics[width=8.5cm]{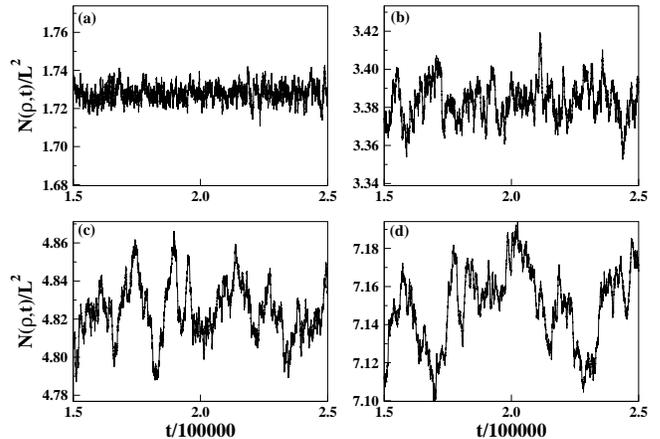}
\end{center}
\caption{
These plots show that as the posting rate $\rho$ approaches the critical
posting rate $\rho_c$ the fluctuation of the average load per site increases
as well as long-range correlation develops. For a system of size $L=128$,
average load $N(\rho,t)/L^2$ is plotted for 100000 time steps
for the posting rates (a) 0.80 (b) 0.90 (c) 0.93 and (d) 0.96 respectively.
}
\end{figure}

      In recent studies on different geometries, that is, on linear chain 
   \cite {Huisinga} and on two-dimensional lattices and on Cayley tree \cite {Arenas}
    a sharp transition from free to congested phase is found for
    routing of packets through shortest paths. In the linear chain the smallest buffer
    causes jamming \cite {Huisinga}. In the Cayley tree the role of the node at the
    top of the hierarchy is crucial for congestion. Also in the case of two-dimensional lattices
    it is observed that if the packet delivery capacity of the nodes is fixed, or 
    independent of the load on the node then congestion occurs above a specific value of the 
    posting rate $\rho$. Traffic flow has also been studied on scale-free networks \cite {Tadic}.

     In this work we try to address this question of dependence (or
   independence) of the network traffic flow on topological features and on the details of the
   dynamic process associated with the generation and flow of traffic. We took a simple
   network of oriented square lattice and select random diffusion along a preferred direction
   as the method of routing of data packets (all of same size) and show that this 
   arrangement could generate the main experimental findings of Internet traffic flow.

\begin{figure}[top]
\begin{center}
\includegraphics[width=7.0cm]{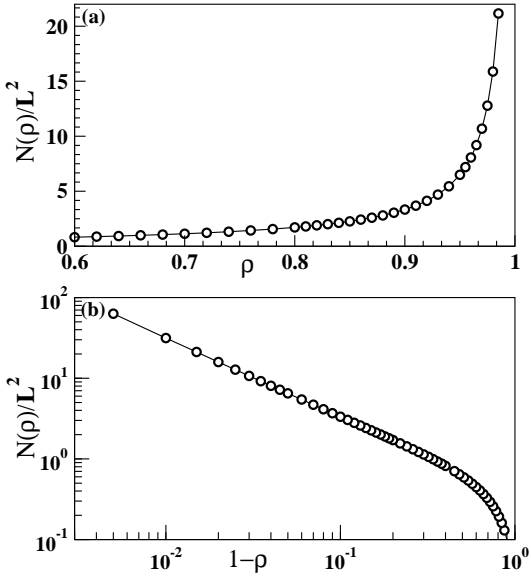}
\end{center}
\caption{
 (a) Plot of the average load per site $N(\rho)/L^2$ with $\rho$ for $L$ =64 
 and for $m$ = 1. The growth of average load diverges at the critical
 point of phase transition at $\rho_c=1$.
 (b) Plot of the average load again but with $1-\rho$ and on a double logarithmic scale,
 the slope gives the value of the exponent $x \approx 0.96$.
}
\end{figure}

\section{The model}

      An oriented square lattice of size $L \times L$ placed on the $x - y$ plane is the
   network in our model: the lattice sites are the nodes and lattice bonds are the links
   of the network. The system has a preferential direction, called `downward' direction, imposed along the $-y$ 
   direction such that packets from every site jump with a positive component 
   along the preferred direction. Every node has two neighboring nodes along the preferred direction 
   which are situated at the lower-left (LL) and lower-right (LR) positions. Data packets are posted at a rate $\rho$ only 
   on nodes of the top row of the lattice at $y=L$. Similarly all nodes on the bottom 
   row at $y=0$ are considered as sinks where packets are delivered and therefore 
   disappear from the system. Though the data packets are distinguishable a packet is 
   delivered at any arbitrary sink. In general each of $L^2$ nodes is a router which 
   receive, store and forward  packets along the preferred direction. There is a limit
   to the forwarding capacity of each node, a node can forward a maximum of $m$ data 
   packets at a time. Each of these data packets are forwarded to LL or LR nodes randomly 
   with equal probability. Each node receives packets from its two upward neighbors, 
   place them in its buffer maintaining a queue of length $q_i(\rho,t)$ and forwards a 
   maximum of $m$ packets at a time from the front of the queue according to the 
   first-in-first-out (FIFO) rule. We further assume that the buffer capacity of 
   each node is infinitely 
   large so that no packet is lost due to filled-up buffer. A single time step during the 
   evolution of the system consists of updating every node of the system for once.

\begin{figure}[top]
\begin{center}
\includegraphics[width=7.0cm]{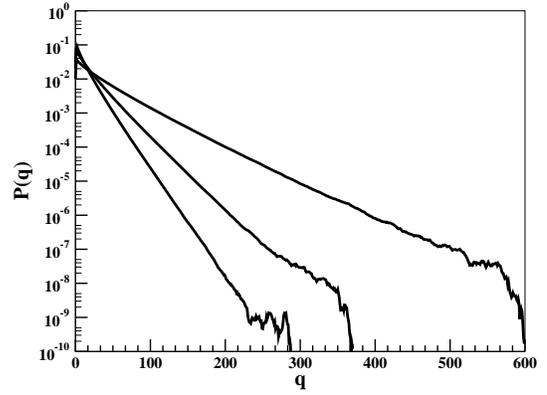}
\end{center}
\caption{
 Queue length distribution for a $64 \times 64$ lattice with $\rho = $
 $0.97, 0.98$ and 0.99. Plots indicate exponential decay.
}
\end{figure}

      The posting rate $\rho$ of data packets is the only control
   parameter of the system. Therefore, at each time step, each node of the top row
   receives a new data packet with a probability $\rho$. The free-flow phase is
   a stationary state where the average fluxes of the inflow and outflow currents 
   of data packets balance. Once a specific value of $m$ is assigned, magnitudes of
   these currents can increase at most to $m$. This implies that the critical
   posting rate $\rho_c$ must be equal to $m$. This is supported numerically
   for a number of values of $m$. In most of our calculations reported below
   we have used $m=1$.

      The total number $N(\rho,t) = \Sigma^{L^2}_{i=1} q_i(\rho,t)$ of data packets
   in the network at time $t$ is called the `load' which fluctuates with time but maintains a steady mean
   value $N(\rho)$ in the stationary free-flow state (Fig. 2 and 3). For a posting rate
   $\rho > \rho_c$ the system switches over to a congested phase and the $N(\rho,t)$
   increases indefinitely. Since packets are moving into the system
   a rate larger than the outflow rate, packets simply pile up in the system
   and no flow balance is attained. The variation of $N(\rho)$ with $\rho$ is
   studied. Different packets take different travel times, 
   to reach their destinations. The probability distribution of these travel times
   is also measured for different $\rho$ values. The nodal queue length distribution
   of the network is studied for different posting rates. 

\section{Results}

      The fluctuation of mean load per node or the mean queue length ${\bar q(\rho,t)} = N(\rho,t)/L^2$
   is observed to have a self-similar fluctuation as shown in Fig. 2 for $\rho=0.96$ and $L=128$. In the Fig. 2(a)
   $N(\rho,t)/L^2$ has been plotted with time $t$ over an interval of length
   $\Delta t = 10^5$. A small boxed region over a time interval of $\Delta t = 10^4$
   from Fig. 2(a) has been zoomed to Fig. 2(b) using a vertical magnification 2.72
   having the same size as of 2(a). Similarly
   a boxed region over a time interval of $\Delta t = 10^3$ from Fig. 2(b) has been zoomed to
   Fig. 2(c) using a vertical magnification 3.48 having the same size as of 2(b). It is evident from the three plots
   that apart from the stochastic noise present in the system the fluctuation of mean load is 
   self-similar.
   
      In addition the fluctuation of mean load per node becomes stronger and more
   correlated when $\rho$ approaches the critical posting rate $\rho_c$.
   For small $\rho$ value the load fluctuates around
  an average value and the fluctuations are also small. It means that the correlation
  time is short, that is state of the system at a certain time step has very little 
  effect on the states of the system a few time steps away. As $\rho \to \rho_c$ the
 fluctuations in the mean load becomes increasingly stronger and it spreads in a wider region indicating
 higher correlation in the system. Near $\rho_c$ a particular state of the
 system naturally influences the states of the system far away
 from it.
   In Fig. 3 we plot $N(\rho,t)/L^2$ for four different posting rates $\rho = 0.80, 90, 0.93$
   and 0.96 over an interval of length $10^5$ time units and for the system size
   $L=128$. The width of fluctuation $w(\rho) = {\langle {\bar q^2(\rho,t)} \rangle - \langle {\bar q(\rho,t)} \rangle^2}$
   increases as posting rate increases and also the fluctuation becomes more and more correlated.

\begin{figure}[top]
\begin{center}
\includegraphics[width=6.5cm]{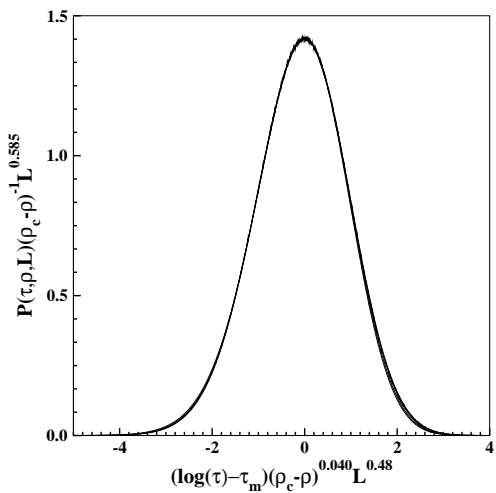}
\end{center}
\caption{
Scaling plot of the data for the distribution of travel times $P(\tau,\rho,L)$
on systems of sizes $L=32$ and 64 for posting rates 0.95, 0.97 and 0.99.
The scaled plot fits nicely with a log-normal function.
}
\end{figure}

   When $\rho$ is very small the 
   number of packet posted per time step is also very small and the system can
   deliver it very quickly to the destination. No queue could be formed on the nodes
   and a packet need not wait in any node while traveling. This is a free-flow phase.
   But when the posting rate $\rho$ increases, slowly queues are formed and a packet 
   had to wait in queues while traveling and this waiting-time started contributing
   to the travel time of a packet. But up to certain $\rho$ value the queue length
   and hence the waiting-time at the nodes fluctuates around an average value, i.e,
   still the average delivery rate of the system and the average posting rate is
   equal and there is
   no growing accumulation of packets in the system. At this stage queues are
   formed on the nodes and average length of the queues increases with $\rho$
   but that average value is not growing with time. But as $\rho \to \rho_c$ posting rate
   becomes equal to the maximum delivery rate (capacity) of the
   system and beyond that there will be growing accumulation of packets in the system
   indicating congestion or jammed phase. 
       
      Though exactly at $\rho=\rho_c=m$ the balance of inflow and outflow fluxes of data packet
   currents is maintained it is observed that no stationary state is attained at this critical point. This
   is because not all nodes of the bottom row at $y=1$ receive exactly one data packet each at every
   time unit, by fluctuation some nodes receive two and some other nodes do not 
   receive any packet at all. Since a node even if received two data packets can deliver
   at most one packet, some packets must have to stay back in the system ultimately leading to a global
   congestion. How the mean load per node increases with time at the non-stationary
   state of $\rho=\rho_c$? It is observed that the variation is parabolic i.e.,
   ${\bar q}(\rho_c,t) \sim t^{1/2}$. However for $\rho > \rho_c$ the growth is 
   linear in time: ${\bar q}(\rho_c,t) \sim t$.
   
      The time averaged load per node $N(\rho)/L^2$ in the free-flow stationary state is calculated for different
   values of the posting rates $\rho$ and are plotted in Fig. 4(a) for $L=64$. For small
   values of $\rho$ the load is small and increases slowly. However when $\rho$
   approaches $\rho_c$ the rate of increase is very fast and diverges. This is seen
   more explicitly in Fig. 4(b) where $N(\rho)/L^2$ is plotted with $\rho_c-\rho$
   on a double logarithmic scale and a straight line is obtained for $\rho$ close
   to $\rho_c$. The slope of the straight line is 0.98, implying that
   the load may vary as:
\begin{equation}
N(\rho)/L^2 \sim (\rho_c-\rho)^{-1}
\end {equation}
   
  Queue length distribution $P(q)$ of the system which is analogous to jam size
 distribution in the highway network gives a better understanding of the
 packet flow scenario. Compared to single-queue theories, here we have many 
 interacting queues in which at each time step a single packet can hop from
 any queue to any of the two neighboring queues. Thus here apart from the
 source nodes all other nodes are placed equivalently. They all have only two 
 neighboring nodes as source of data packets. In Fig. 5 we show the plot of $P(q)$ vs. $q$
 for $\rho = 0.97, 0.98$ and 0.99 for $L=64$ on a 
 semi-log scale indicating that the intermediate region of the distribution follow
 the exponentially decaying distribution $\exp(-q/q_o(\rho))$ in general. It has been observed 
 that the dependence $q_o(\rho) \sim (\rho_c-\rho)^{-1}$ is followed very nicely.
 The average queue length $\langle q(\rho,y) \rangle$ is also measured as a function of 
 the $y$ co-ordinate and found to vary as $(\rho_c-\rho)^{-1}$ independent of $y$ except over
 a small region near the top level.

      The travel time $\tau$ of a data packet is defined as the time spent by the packet
   in the system which is obviously the difference between the delivery and posting times.
   Each data packet is given a label and with each node a queue list is associated containing the
   labels of the data packets in this queue. If the queue length at a node is greater
   than $m$ then first $m$ packets from the front of the queue are deleted and the queue is
   shifted $m$ location to the front. Each of these $m$ packets are then randomly routed
   to one of the LL or LR node. Such a packet is placed at the end of the queue in the new node 
   by the FIFO rule.
   Intuitively it is easy to understand that for very low posting rates $\rho \to 0$
   every data packet makes a hop at every time instant and therefore
   all $\tau$ values are same and equal to $L$. Consequently the $P(\tau,\rho, L)_{\rho \to 0} = \delta (\tau-L)$.
   However as $\rho$ increases the queue lengths become larger, as a result travel time
   increases and its distribution gains a finite width. We have studied in detail the
   distribution of travel times and its dependence on $\rho$ and $L$ and 
   $P(\tau,\rho,L)$ is observed to follow a combined scaling form over $\rho$ and $L$ as:
\begin {equation}
   P(\tau,\rho,L)(\rho_c-\rho)^{-1}L^{0.585} \sim {\cal G} [(\log(\tau)-\tau_m)(\rho_c-\rho)^{0.04}L^{0.48}]
\end {equation}
   The scaling function ${\cal G}(x)$ is seen to follow a log-normal function like:
\begin {equation}
 {\cal G}(x)=\frac {1}{x\sigma \sqrt{2\pi}}\exp{(-\frac{(\ln x)^2}{2\sigma ^2})}
\end {equation}

\begin{figure}[top]
\begin{center}
\includegraphics[width=7.5cm]{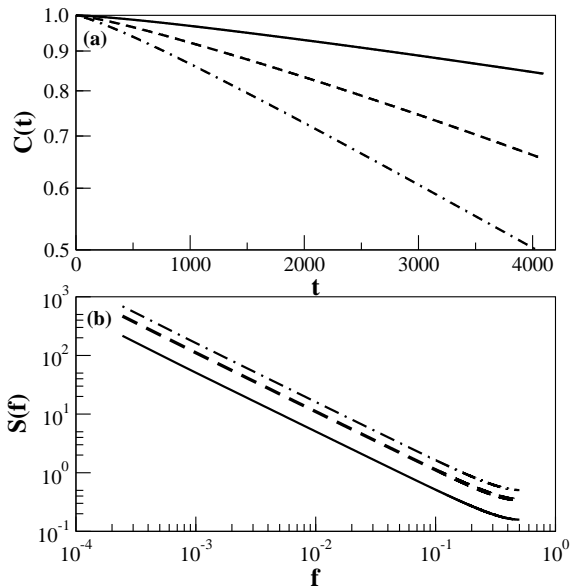}
\end{center}
\caption{ 
 Plot of (a) correlation function $C(t)$ vs. $t$ on a semi-log scale and
 (b) power spectrum $S(f)$ vs. $f$ on a double log scale for the posting
 rates: $\rho = $ 0.96 (dot-dashed), 0.97 (dashed) and 0.98 (solid) lines
 for a system size $L$ = 64.
}
\end{figure}

\section
{
Power-spectral analysis of network time-series
}
      The fluctuation of mean queue length per node $\bar q(\rho,t)$ with time depends on the
   posting rate. In the free-flow stationary state the autocorrelation function of $\bar q(\rho,t)$
   is defined as:
\begin {equation}
C(\rho,t) = \frac { \langle {\bar q(\rho,t')\bar q(\rho,t+t')} \rangle - \langle {\bar q(\rho,t')} \rangle^2}
      {\langle {\bar q^2(\rho,t')} \rangle - \langle {\bar q(\rho,t')} \rangle^2}
\end {equation}
    Fourier transform of the  autocorrelation function $C(\rho,t)$ is known as the spectral density or
   power spectrum $S(f)$ defined as
\begin {equation}
 S(\rho,f) = \int^{\infty}_{-\infty} e^{-ift}C(\rho,t)dt
\end {equation}
   For a time series which has no temporal correlation, plot of $S(f)$ against $f$ is 
   independent of $f$. For some other time data series, the power spectrum may vary as a power law:
   $S(f) \sim f^{-\phi}$. In this case
   the spectral exponent $\phi$ characterizes
   the nature of persistence, $\phi = 2$ indicates zero correlation associated with Brownian 
   motion, $\phi > 2$ indicates positive correlation and persistence and $\phi < 2$
   represents negative correlation and anti-persistence.
 
      The autocorrelation $C(\rho,t)$ function of the mean queue length in the stationary state
   is plotted in Fig. 7(a) on a semi-log scale up to $t$=8192 for three values of 
   the posting rate $\rho = 0.96, 0.97$ and 0.98 calculated on a system size $L=64$.
   Fourier transformations of these correlation functions are done using `{\it xmgrace}' and the
   power spectrum $S(f)$ is plotted with $f$ on a double logarithmic scale in Fig. 7(b) for all
   three values of posting rates. The intermediate regimes of the curves are quite straight 
   indicating a power law variation of the power spectrum: $S(f) \sim f^{-\phi(\rho)}$.
   From Fig. 7(b) we measure the slopes are all nearly same and $\phi(\rho) \approx 1$
   indicating $1/f$ noise near critical posting rate $\rho_c$
   irrespective of the precise value of the posting rate.
   
      Similar autocorrelation functions and associated power spectrum are also calculated for the
   fluctuation of the length of a single queue $q_i(\rho,t)$. The power spectrum is also observed
   to follow a power law with the spectral exponent value nearly equal to one.

      Finally the un-directed version of this problem has also been studied on the square lattice.
  In this case the data packets are posted at the nodes on the middle row $y=L/2$
  and are delivered at the nodes on the top row at $y=L$ and the bottom row $y=1$ with
  periodic boundary condition applied along the $x$-axis. Each packet executes a simple
  non-interacting random walk i.e., for each step it selects one of the four neighboring
  nodes randomly with uniform probability and jumps to that site. As before a similar
  phase transition is observed from a free-flow state to a congested phase at a specific
  posting rate $\rho_c$. However unlike the previous model, the critical rate $\rho_c \to 0$
  as $1/L$. This is because a large number of packets simply pile up at the middle line
  for all values of the posting rates greater than $1/L$. Travel times of packets
  again follow log-normal distributions and power spectrum also follows a similar power
  law.
    
\section {Conclusion }

    The properties of traffic of the flow of data packets on a model network of oriented square
 lattice with random routing scheme largely consistent with the real-world
 Internet and vehicular traffic flow behaviors. First, it describes the transition from free
 phase to congested phase with the increase of density of packets separated by a
 well-defined critical posting rate. Second, it produces the self-similar nature
 of network workload time series. Third, the long-tailed (log-normal) nature of
 the travel time distribution is reproduced. Fourth, the power spectral
 analysis shows $1/f$ type noise confirming long-ranged correlation in the network load time series 
 near criticality. 

  We thank B.Tadic for useful discussion. GM thankfully acknowledged facilities 
  at S. N. Bose National Centre for Basic Sciences.  

\leftline {Electronic Address: manna@bose.res.in}

\end {document}